# Effect of size and aspect ratio on structural parameters and evidence of shape transition in zinc oxide nanostructures

**Manoranjan Ghosh,\*  Debjani Karmakar, S. C. Gadkari, and S. K. Gupta**

Technical Physics Division, Bhabha Atomic Research Centre, Mumbai-400085, India

**S. Basu, S.N. Jha, and D. Bhattacharyya**

Applied Spectroscopy Division, Bhabha Atomic Research Centre, Mumbai-400085, India

**Abstract**

Dependence of structural parameters on the size of nanoparticles is a topic of general interest where the effect of shape is often neglected. We report a comprehensive study on size dependent structural parameters of ZnO nanostructures (NS) having a wide range of aspect ratios (length/diameter). It reveals that with increase in size, ZnO NS undergo a shape transition from spherical to rod like morphology that induces sudden change in internal parameter (u) which represents relative position of two hexagonal close-packed sublattices. The change in u introduces concomitant changes in anion-cation (Zn-O) bond lengths as well as bond angles and thereby bears a linear dependence with the aspect ratio (AR). Further, the unit cell volume and microstrain decrease with increase in particle size and show a drastic reduction when flat crystal faces begin to appear at the spherical surface (AR~1.3). The significant change in structural parameters associated with the shape transition arises due to surface dipole induced electrostatic relaxation that may be further influenced by interaction with the ambient gases as evidenced from the Extended X-ray Absorption Fine Structure (EXAFS) measurement. The present study addresses the underlying reasons of shape induced change in structural and electronic properties of ZnO NS.

## 1. Introduction

There remain continued interest on ZnO nanostructures of different shapes and sizes exhibiting promising properties for application in short wavelength optoelectronic devices and sensors. Nanorods (NRs) of ZnO often show vastly different properties than nanospheres [1–5]. Specific crystal planes interact differently with the ambient gases and can be controlled by changing the aerial ratio of polar to non-polar faces of ZnO NRs [6]. A thorough investigation on the subtle changes in structural parameters due to change in size and shape of ZnO nanostructures (NS) may fill the gap in this issue. Size dependence of structural parameters in nanoparticles (NPs) is not straight forward and many case-specific reports are available. In metal NPs, lattice parameters as well as cell volume increases with increasing particle size that has been explained by a model based on elastic properties [7,8]. The trend is different in metal oxide NPs, where the unit cell volume expands due to size reduction [9,10] and the effect is understood in terms of reduced electrostatic forces due to surface dipoles. In rare events contraction of lattice due to size reduction is also seen, which has been attributed to a positive pressure due to surface hydration [11].

Although, ZnO NS of various shapes and sizes are very common, there are limited reports available on the size dependence of lattice parameters of ZnO NPs. The lattice constant $c$ of ZnO NRs decreases when grows by length towards $c$-axis ([0001] direction) [12]. However, in most of the above cases, shape of the NS is not given its due importance in the analysis of the results. Independent investigation on shape and size dependent properties requires NS of different shapes and sizes having same size and shapes respectively. In case of ZnO, it is observed that the growth in size is generally associated with a change in shape. Nanosphere when grows in size produces flat faces by relaxation of built in strain and the shape of a nanocrystal deviates from spherical one [ie., aspect ratio (AR) > 1]. Therefore, synthesis of NS of different sizes (shapes) having same shapes (sizes) is difficult. In this article a large number of samples with wide range of size and shapes have been considered. An appropriate selection of samples and in-depth analysis of the results reveal that the effect of size and shape are indeed distinguishable.

The ZnO NS may act as a host for electrostatic charge at the surface which also depends on ambient environment [13]. The large electrostatic field created by this surface charge should not be treated as the confinement effect due to the size reduction [14]. A recent ab initio calculation by Dag and co-workers reveals large side surface dipole moment of ZnO NR [1]. It may create a non-uniform distribution of surface charge that will induce a change in the Zn-O bond length specific to a crystal plane. The influence of ambient environment which is a key factor in determining various surface related properties needs to be taken into account. The crystal faces which grows faster (slower) than others will have a larger (smaller) area of contact per unit volume with the ambient environment. Thus the impact of interaction with the ambient will be different for different faces of crystal having anisotropic growth rate. For one-dimensional (1D) ZnO NRs that grows preferentially along the [002] direction, close packing of (002) planes is expected [12]. Since $c$ and $a$ axes are the natural growth directions for rutile and anatase phases of $TiO_2$ respectively, variations of the corresponding lattice constants with respect to particle size are also reported to be different [3,15]. Further, electrostatic interaction with ambient environment and the crystal structure were claimed to be responsible for the observed size dependence of lattice parameter in $CeO_2$ and $BaTiO_3$ NPs [16]. Thus the variation in structural parameters due to a change in size and shapes of ZnO NS depend on intrinsic electrostatic relaxation due to surface dipole moments and the contact area between nanomaterials and the ambient environment which is a function of AR.

In this article, the size dependence of structural parameters of ZnO NS has been investigated which was hitherto not been addressed in details. It is seen that growth in size of the NS is associated with a change in shape from spherical to rod like morphology having flat faces. The apparent change in shape is reflected by a sudden transition in the structural parameters specific to particular crystal planes. With the help of ab-initio first principles techniques, we have investigated the structural effects of size-confinement in vacuum for ZnO clusters of different ARs. The results show a qualitative resemblance with the XRD analysis. However, for small spherical particles having high surface to volume ratio, influence of ambient gases predominates which has uniform effect on various structural parameters. To establish the role of ambient gases, in-situ Extended X-ray Absorption Fine Structure (EXAFS) measurements have been carried out in the presence of different gases using synchrotron radiation. Further, the experimental findings are partially verified with the results extracted after relaxation of ZnO clusters having different aspect ratios in the framework of ab-initio density functional theory (DFT).

## 2. Experimental details and method of analysis

ZnO NPs of a wide range of size and AR have been prepared by chemical route. Three types of nanomaterials having wide range of size and AR have been considered in this study. The size and AR of all the samples investigated are plotted in Fig. 1 and shown in ref. 2, 17, 20, 34. Type I samples have size below 15 nm and AR in the range of 1-1.3 (Fig. 1). These samples are highly mono-disperse having very small variation in diameter (ie., $\Delta d/d <$ 10%). The general methodology used in synthesizing ZnO NS using zinc acetate dihydrate as precursor is described elsewhere [18,34]. Briefly, the NaOH solution in ethanol is slowly added into the solution of zinc acetate dihydrate in ethanol kept at 60-75 $^0$C. This method produces nearly spherical NPs in the size range of 5-15 nm. To achieve a greater control over their size and morphology, the above mentioned reaction has been performed under high pressure and temperature in an autoclave as described elsewhere [2,19]. Samples prepared by this technique (Type II samples) have a wide range of size (20-100 nm) and AR (1.4-2). Different precursors have been used to grow the ZnO NRs of very high AR ranging from 5-12 (Type III samples). These NRs can be grown on substrate as described elsewhere [20].

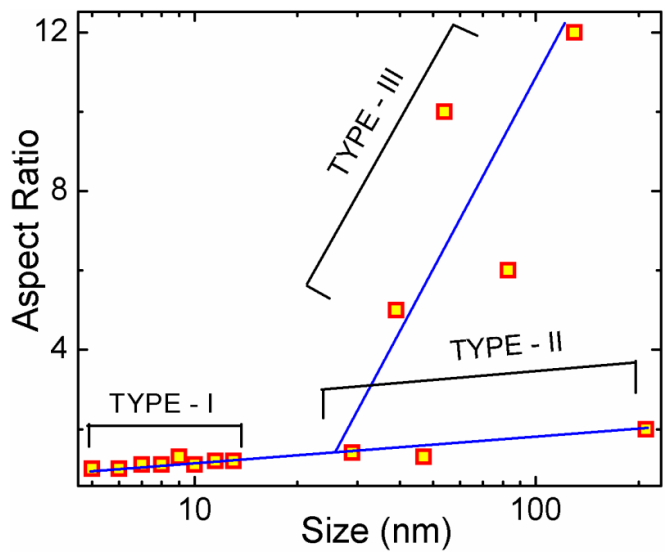

**Fig. 1.** (color online) Aspect ratio (AR) vs. size plot of nearly spherical (Type I), rods (Type II) and long rods (Type III) of ZnO nanostructures exhibit tilted Y-shape characteristic.

Size and morphology of the NPs have been determined by complimentary techniques like XRD, TEM and SEM and shown in ref. 17. As depicted in Fig. 1, the plot of AR vs. particle size represents a tilted Y shape which has a great significance. There are three types of NS such as nearly spherical particles of size below 20 nm (Type I), rods of diameter~30-200 nm and AR between 1.5-2 (Type II), long rods of diameter~40-150 nm and AR 5-12 (Type III). The AR of majority of the samples increases steadily with size (Type I and II). But a few of them have very high AR that shows steep rise with the particle size (Type III). Because of this particular nature of the curve, it is possible to distinguish the effect of size and shape. Size dependence of certain quantity which represents tilted Y like shape clearly indicates a definite trend with the AR. In our analysis the size determined from XRD results has been considered because all other structural parameters have been found from the same. For obvious reason, AR of the NS has been found from TEM and SEM images.

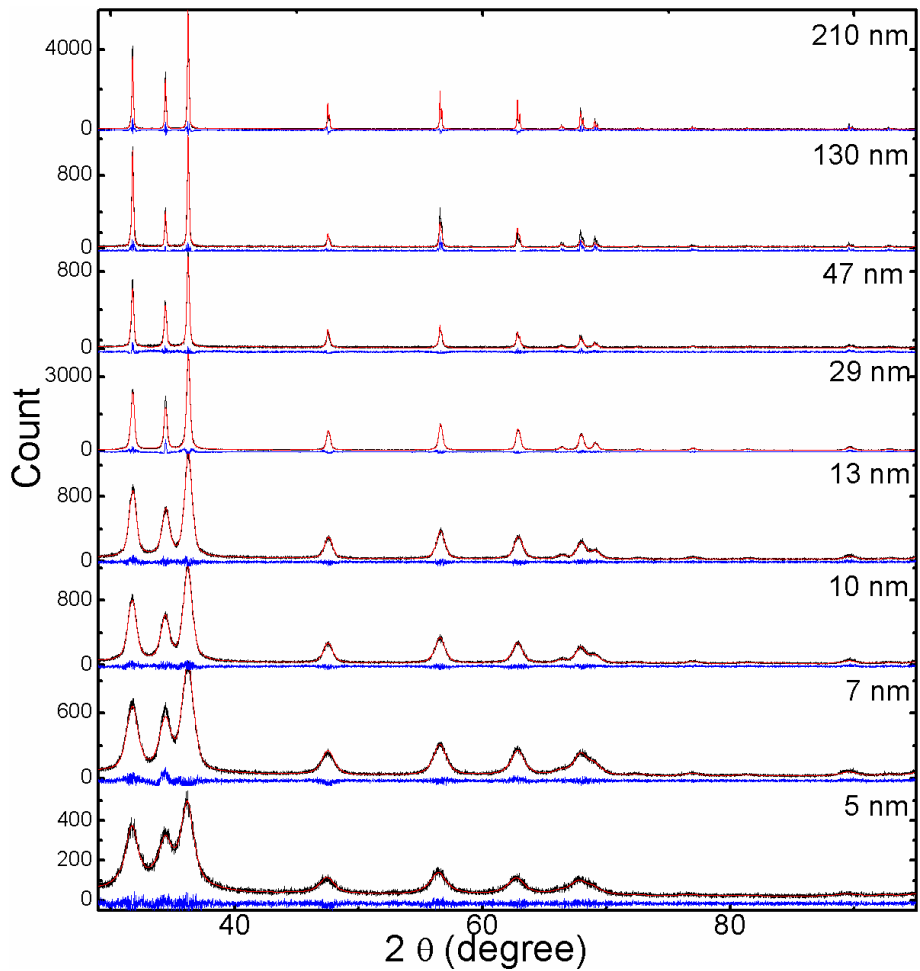

**Fig. 2.** (color online) Full-Prof Fitting of X-ray Diffraction data from ZnO NPs of various shapes and sizes.

All samples were investigated by Philips Xpert Pro X-ray diffractometer and subsequently analyzed to extract the structural parameters. Williamson-Hall method [26] has been used to determine the particle size and strain from the XRD data (described in ref. 17). Size and morphology of the smaller samples were determined by transmission electron microscope (JEOL HRTEM) measurements. Scanning electron microscope results are considered for samples having large dimension and high AR. The EXAFS measurements with synchrotron radiation at the Zn K-edge (9659 eV) in different gaseous atmospheres were carried out at the dispersive EXAFS beamline (BL-8) at the INDUS-2 Synchrotron Source (2.5 GeV, 100 mA) at the Raja Ramanna Centre for Advanced Technology (RRCAT), Indore, India [21,23]. Technical details of the EXAFS measurement facility are described in ref. 17.

The structural parameters of the NPs have been determined from the Rietveld fitting of the XRD data by using FullProf suite [27]. The fitted data along with the experimental one is shown in Fig. 2 for select samples. The method of fitting is described in ref. 17 and 27. ZnO nanomaterials considered here show wurtzite structure (S.G. P63mc). Size variation of the samples is clearly visible from the change in FWHM of the X-ray reflections. In this full profile fitting procedure, pseudo-voigt peak shape was chosen. Important parameters that were refined includes background, lattice parameters, FWHM parameters, shape and asymmetry parameter and internal parameter u. Residues are shown at the bottom of each XRD data presented.

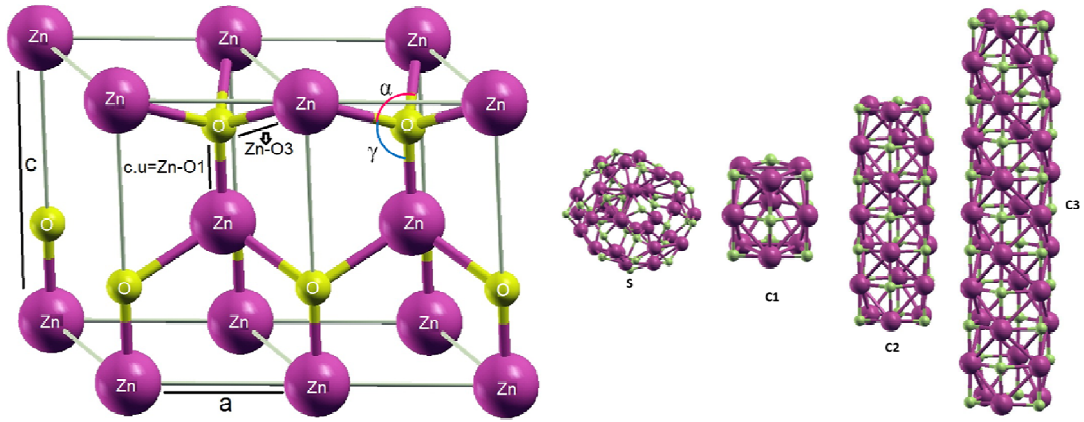

**Fig. 3.** (color online) Important structural parameters of hexagonal ZnO (left). The theoretical cluster systems S, C1, C2 and C3 having ARs 1, 1.7, 4.3 and 7.1 respectively.

## 3. Theoretical Calculations

Structural relaxation-induced modifications of ZnO clusters with different aspect ratios are carried out in the framework of ab-initio density functional theory (DFT). We have used plane-wave basis set along with the projector augmented wave (PAW) method as implemented in the VASP code [24,25]. PAW potentials with 12 valence electrons ($3d^{10}4s^2$) for Zn and 6 ($2p^4 2s^2$) for O have been used. The clusters are placed inside a large cell, so as to avoid overlapping of periodic images. The plane wave energy cut off was taken to be 600 eV. Single *k* point (gamma point) was used for this calculation. The atomic positions and the lattice parameters are relaxed by using conjugate gradient algorithm into their instantaneous ground-state such that the Hellman-Feynman forces on each atom are minimized with a tolerance value of 0.001 eV/A. We have carried out total energy calculations within local density approximation (LDA) for four clusters with aspect ratios approximately close to 1, 1.7, 4.3 and 7.1, as presented in Fig. 3. We hereby denote these four clusters as S, C1, C2 and C3 respectively. After ionic relaxation in vacuum, these four systems reveal interesting changes in structural details while compared to those of bulk single-crystalline ZnO. The box-size in vacuum is chosen in such a way that a minimum spacer of 20 Å is there from all directions of the cluster.

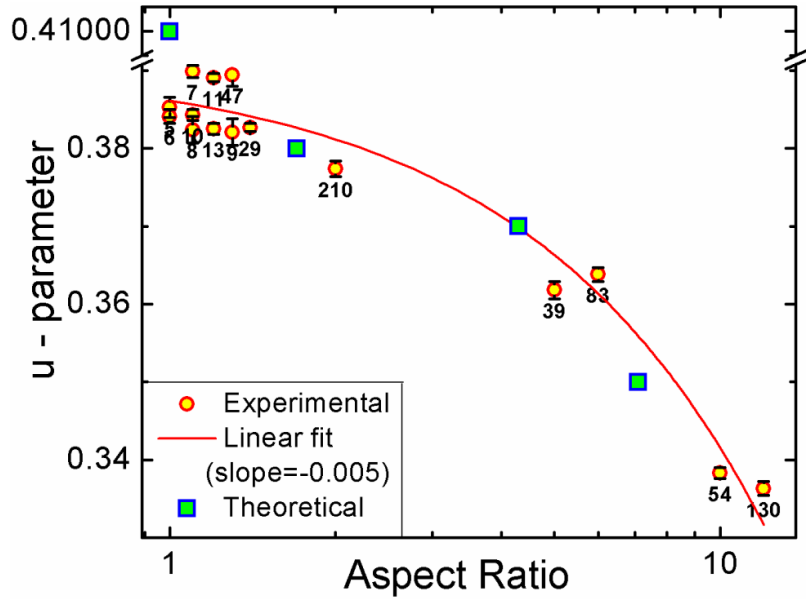

**Fig. 4.** (color online) Internal parameter (u) exhibit approximately linear dependence with the aspect ratio shown in logarithmic scale. Corresponding sizes are indicated.

## 4. Results and discussions

*4.1. Correlation of internal parameter (u) with aspect ratio (AR) and variation of Zn-O bond lengths*

ZnO has a hexagonal unit cell [Fig. 3] with two lattice parameters *a* and *c* in the ratio ideally can be 1.633 [c/a = (8/3)0.5] [28]. Lattice parameters *a* and *c* represent the edge length of the basal plane hexagon and the axial height of the unit cell perpendicular to the basal plane respectively. The wurtzite structure of ZnO is composed of two inter-penetrating hexagonal close packed (hcp) sub-lattices containing four atoms per unit cell of one type. Every atom of one kind is surrounded by four atoms of the other kind placed at the corner of a tetrahedron or vice versa. These sub-lattices are displaced with respect to each other along the *c*-axis by the amount *u*, known as internal parameter. The fractional atomic coordinates of oxygen (1/3, 2/3, *u*) in ZnO (S.G. P63mc) is expressed in terms of *u* which is also linked with anion-cation bond length parallel to the c-axis (Zn-O1, N=1 is the co-ordination number) and the lattice parameter c by the relation c.u=Zn-O1 [Fig. 3]. Reduction of *u* is possible when Zn-O1 decreases or *c* increases. It can also induce change in Zn-O3 (N=3) bond perpendicular to c-axis and bond angles α and γ.

As shown in Fig. 4, *u* decreases as AR increases exhibiting approximately linear dependence. So, the functional dependence of other structural parameters on *u* and AR shows similar characteristics. Similar to Fig. 1, size dependence of the *u* parameter exhibits tilted Y shaped geometry (not shown). There is no significant change in *u* with size for spherical NPs but shows sudden reduction in case of NRs of high AR indicating a definite trend with the variation of AR. The calculated *u*-parameter of clusters having different AR qualitatively supports the experimental behaviour as plotted in Fig. 4. Starting from S to C1, C2 and C3, the u-parameter evolves as 0.4, 0.38, 0.37 and 0.35 respectively.

The dependence of Zn-O1 and Zn-O3 on both u and AR is shown in Fig. 5. These bond lengths are linear functions of *u* but exhibit minor deviation from linearity with AR for NRs (Type III). There is stretching of Zn-O1 along *c*-axis [(0001) face] and contraction of the Zn-O3 towards perpendicular to *c*-axis (side wall) of ZnO NR with the increase in *u*. Or in other words Zn-O1 decreases and Zn-O3 increases as the AR increases. Slopes of the variation of Zn-O1 and Zn-O3 vs. *u* are 5.22 and −1.79 as determined from the linear fitting of the data points. The change in Zn-O1 is significantly larger (15-20 %) compared to Zn-O3 showing maximum change below 5%. Theoretical evolution of Zn-O1 and Zn-O3 with AR and *u* also has a qualitative resemblance with the experimental data as plotted in Fig. 5. For the bulk-system, Zn-O1 and Zn-O3 are found to be 1.77 Å and 2.046 Å respectively which is nearly equal to the experimentally obtained results for long NRs (Type III). The mismatch of the slope with the experimental results, may appear due to several reasons, viz., (1) there is large gap in size between theoretical clusters (0.1-2 nm) and samples obtained experimentally (>5 nm); (2) the experimental samples may have a size-distribution, whereas the theoretical system is of a definite size; (3) moreover, the environment effects and presence of vacancies, as is expected to be present for experimental systems, are absent in case of theoretical studies.

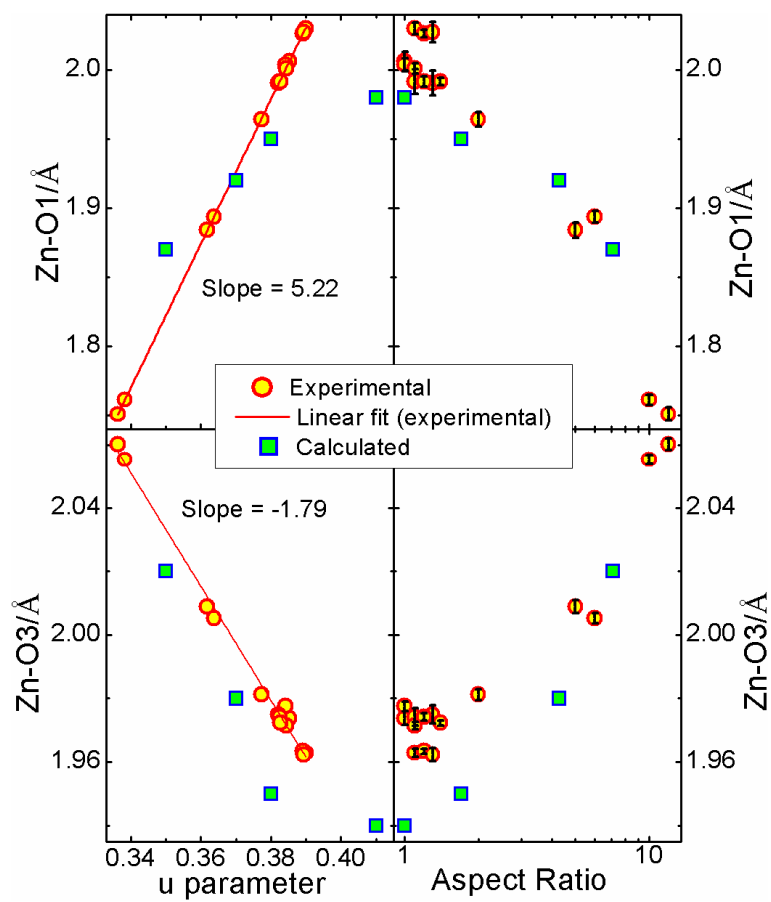

**Fig. 5.** (color online) Anion-cation bond lengths (top) along c-axis [Zn-O1 (co-ordination number N=1)] and (bottom) off c-axis [Zn-O3 (N=3)] as a function of internal parameter (*u*) (left) and aspect ratio (right). X-axis of right panel is shown in logarithmic scale.

Zn-O1 and Zn-O3 exhibit tilted Y shape variation with particle size as expected from their linear dependence on u (Fig. 6). These bond lengths show a sudden jump when shape of the NS changes. A significant enhancement in Zn-O3 and reduction in Zn-O1 can be seen for long NRs having AR>5. But for smaller NPs both Zn-O1 and Zn-O3 marginally reduce when size increase. This apparent unusual variation of the bond lengths vs. size will be discussed in the subsequent sections. There are instances when a clearly faceted NS having disk like morphology is obtained. These NS show low AR (~1.3) having diameter ~50 nm and thickness~70 nm. The *u* parameter of these NS is very high along with Zn-O1 that reduces Zn-O3.

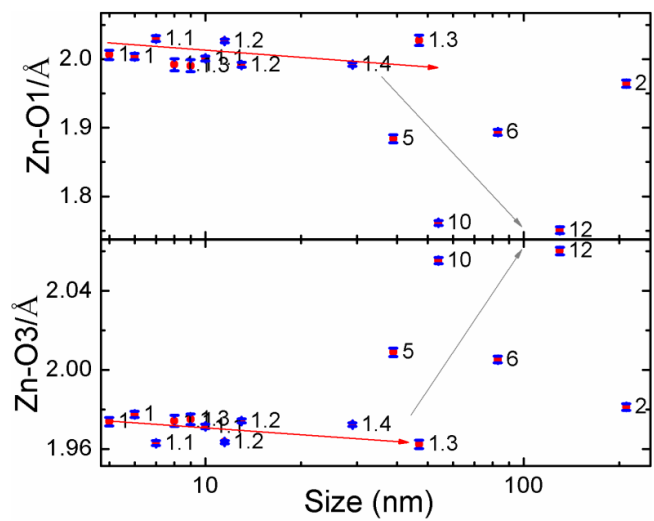

**Fig. 6.** (color online) Size dependence of Zn-O bond lengths show tilted Y-shape characteristic. AR is indicated near data points.

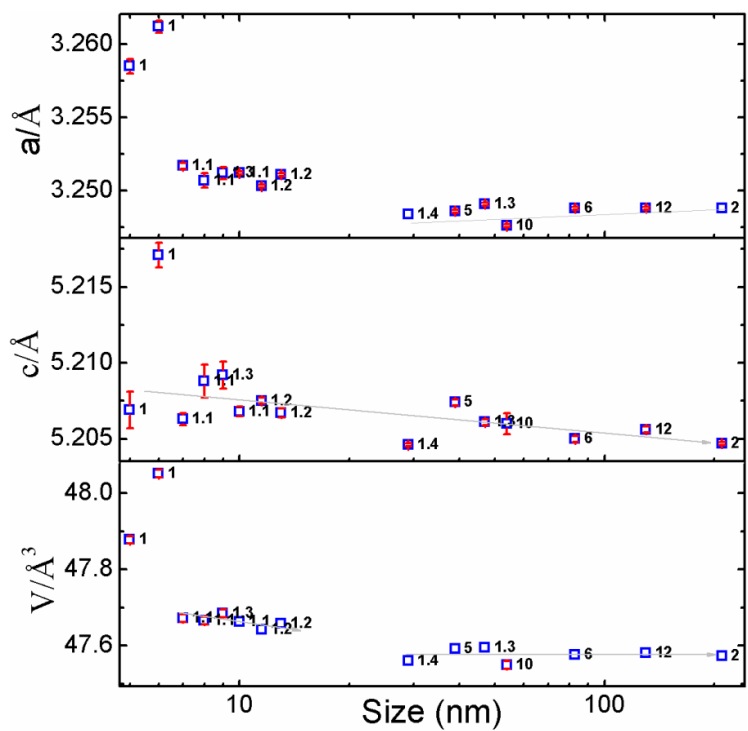

**Fig. 7.** (color online) Lattice parameters *a*, *c*, and *V* vs. size. AR is indicated near data points.

*4.2. Size dependence of lattice constants and shape transition in ZnO nanostructures*

At this point it is important to check the variation of the cell parameters *a*, *c* and *V* with size and AR. As shown in Fig. 7, size dependence of the cell parameter can be divided in to two regions. For AR < 1.3, *a*, *c* and *V* decrease with increase in size. It is consistent with the observed reduction in both Zn-O1 and Zn-O3 for smaller NS (AR<1.3) as seen from Fig. 6. For higher AR (>1.3) *c* decreases but *a* slightly increases with increase in size irrespective of AR. Reduction in Zn-O1 and increase in Zn-O3 for bigger NRs (Fig. 6) corroborate the contraction in *c*, and enhancement in *a* respectively. Therefore the variation of *c* and *a* with size is consistent with the variation of Zn-O1 and Zn-O3 separately for smaller (Type I) and larger NS (Type II and III). But there is a sharp increase in Zn-O3 despite a reduction in lattice constant *a*, when shape changes from spherical (AR<1.3) to rod like morphology (AR>1.3). Such unusual size dependence is evident in case of other structural parameters as well. Zn-O3 first decreases (for Type I) and then increases with AR (Fig. 5) exhibiting a changeover at AR~1.3. The variation of Zn-O3 with size before and after the shape transition is entirely different. Although *u* and Zn-O1 show monotonic reduction for the full range of size but change in shape accelerate the rate of reduction (Fig. 6). Careful observation of Fig. 7 reveals that the variation of *a* and *V* with AR exhibit a clear transition point (AR=1.3) which arises due to change in shape (AR). Thus Zn-O3 suddenly increases despite *a* falls along with reduction in Zn-O1 due to the shape transition. It is possible only when there is sudden change in bond angles γ and α (top inset of Fig. 8). It should be mentioned that the variation of α and γ is complimentary to each other. With size, both these angles exhibit tilted Y-shape variation but their dependence is opposite to each other. As expected, γ linearly decreases with u (data not shown) but α increases with similar magnitude of slope. Therefore apparent change in shape from spherical to rod like morphology is manifested by sudden change in γ and α which stretches Zn-O3 but reduces *a*.

It should be mentioned that theoretically calculated γ of clusters (~90$^0$) is significantly different from that of bulk (~104.68$^0$). There is noticeable change in ionic coordination due to surface reconstruction of theoretical clusters without defects. For cylindrical clusters, the surface coordination changes from tetrahedral to a near-octahedral one, where the off-axial Zn-O-Zn bond-angle (γ) comes very close to 90$^0$. In addition, for the spherical clusters, there is tendency of void formation at the surface, resulting in a large change of surface cation-anion coordination. As a result, γ values for the clusters are significantly lower than the experimental results despite showing similar trend. However, α values for clusters S ($113.02°$), C1 ($112.06°$), C2 ($111.39°$) and C3 ($108.42°$) are close to experimental values (top inset of Fig. 8). Thus, the theoretical investigation implies that even without environmental effects, small clusters of ZnO shows shape-dependent structural modifications which has a good qualitative resemblance with the experimental findings.

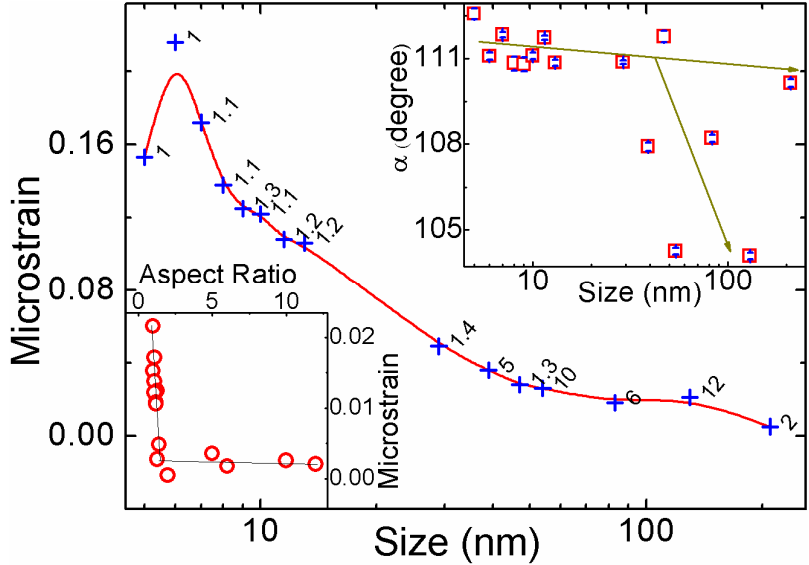

**FIG. 8.** (color online) Micro-strain as a function of particle size and aspect ratio (inset bottom). Variation of bond angle between two Zn-O3 bond ($\alpha$) with size (inset top).

The approach of the cell parameters to the standard bulk value as the size increases is consistent with that obtained after re-crystallization of ZnO film by rapid thermal annealing [29]. In agreement with the prediction of ref. 29, microstrain of the NPs decreases due to increase in size (Fig. 8). When spherical NPs (AR = 1) grow in size, strain builds up. As soon as the shape deviates from the spherical one (AR>1), flat faces appear and the built up strain relaxes (ref. 2). Due to this strain relaxation process, nanomaterials obtain a stable atomic configuration by minimizing the cell volume. If suitable condition prevails, further growth occurs and diameter as well as length of the nanomaterial increases. At this stage not much change in the microstrain is observed and the change in AR has little effect on structural parameters. A clearer picture of the effect of shape can be seen in the microstrain vs. AR plot (bottom inset of Fig. 8). A drastic fall in microstrain is observed when AR of smaller NPs changes from 1 to 1.3. There is minor change in microstrain for AR>1.5. It reiterates that maximum change in properties is expected at the point of shape transition ie., when flat crystal faces begin to appear at the spherical surface of a NP. During this process atoms in the surface layer undergo change in positions that may be responsible for large surface dipole moment in NRs [1] and major changes in electronic properties [2].

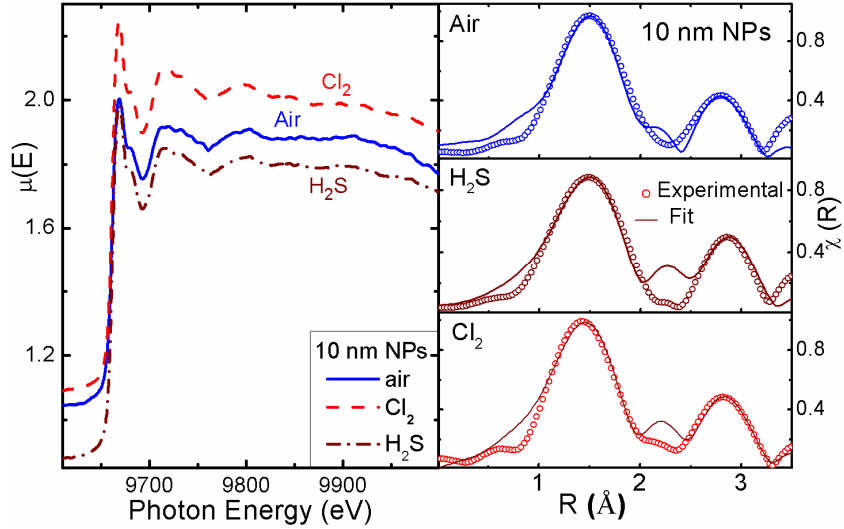

**Fig. 9.** (color online) Experimental EXAFS [μ(E) vs E ] and absorption function [χ(R)] of 10 nm NPs after the exposure of different gases.

*4.3. Origin of observed variation in the structural parameters*

The expansion of lattice due to size reduction in metal oxides reported in the past has been commonly explained in terms of confinement effect, [9] reduced ionic valencies, [15,16] electrostatic relaxation [30] and interactions among the surface dipoles [10]. We also propose that electrostatic relaxation induced by surface dipole moment in ZnO NS is primarily responsible for the observed variation in structural parameters. The *u* parameter of wurtzite ZnO is known to increase due to application of hydrostatic pressure [31]. Therefore increase in *u* with decrease in AR [with increased contribution of (0001) face] can be understood in terms of positive (inward) pressure applied by surface dipoles along c-axis. The observed increase in Zn-O3 and decrease in Zn-O1 for large NRs can be understood in terms of repulsive and attractive force exerted by surface dipole moment perpendicular to side wall and (0001) face respectively as predicted by Dag et al [1]. The shape transition from sphere to rod enhances the bond angle γ in large NRs associated with contraction (stretching) of Zn-O1 (Zn-O3) which brings tetrahedrally bonded O atom closer to each other. The modified charge distribution after this reorientation of the atomic positions may create strong repulsive and weak attractive dipole moment at the side wall and (0001) face of NRs respectively. The contraction in *c* with increase in size is consistent with the observed reduction in Zn-O1. There is a sudden reduction in *a* due to rearrangement of atomic positions in the process of shape transition. After the shape transition *a* increases with size which is consistent with the increase in Zn-O3 and large surface dipole moment predicted for NRs having high AR. Due to opposite variations in *c* and *a* after shape transition, the cell volume remain almost unchanged. This finding may resolve many contradictory variation in *a* and *c* reported for ZnO related materials [32,33]. For NRs having large diameter, influence of surface dipoles and surface energy on the bulk of the crystal is negligible and dependence of lattice parameters on aspect ratio and size diminishes.

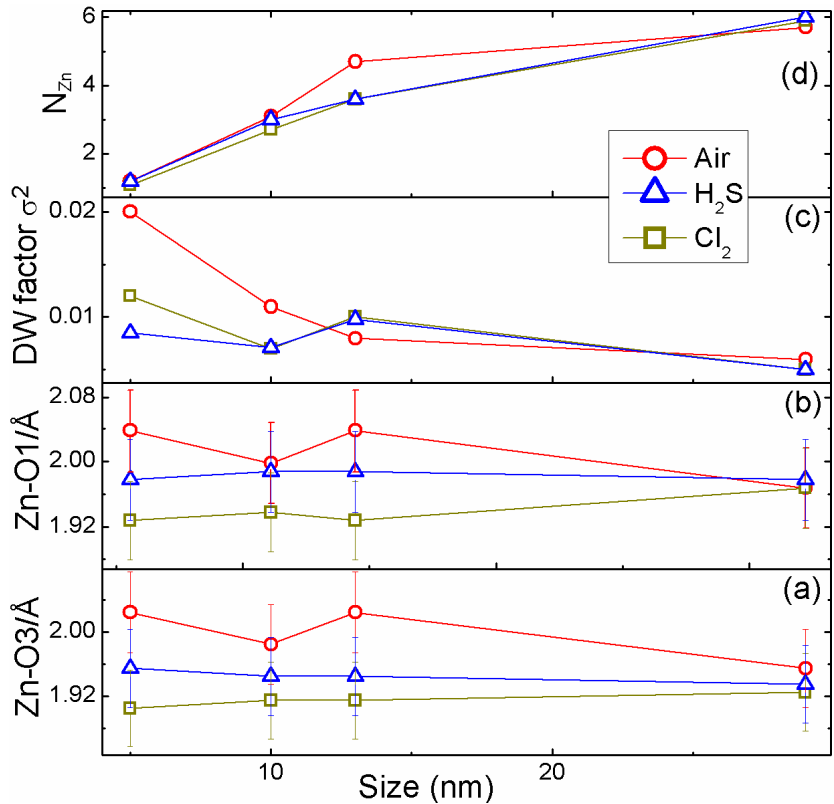

**Fig. 10.** (color online) Variation of (a) Zn-O3, (b) Zn-O1 (c) Debye Waller factor and (d) Zn coordination number of the Zn–Zn shell with particle size when exposed to different gases.

For small spherical NPs both Zn-O1 and Zn-O3 increase marginally as size decrease which is consistent with the expansion in *a*, *c* and V. Also the *u* parameter is high and increases slowly with decrease in particle size. It indicates that surface of smaller NPs irrespective of the direction of crystal planes experience repulsive force (negative pressure). This prompts us to conclude that adsorbed ambient gases have significant effect on the structural parameters of small NPs. To establish the role of environment on the local structural parameters of ZnO NPs, in-situ EXAFS investigation has been performed under different ambient conditions. Fig. 9 (left) shows representative experimental EXAFS [µ(E) vs. (E)] spectra of a Type-I sample (diameter ~10 nm, AR~1) at Zn K-edge for various gaseous environments. In order to take care of the oscillations in the absorption spectra, the energy dependent absorption coefficient µ(E) has been converted to absorption function [χ(R)] in terms of real distance from the centre R of the absorbing atom (see ref. 17 for the method of analysis). The best fit of experimental spectra of ZnO NPs (diameter ~10 nm) in air, $Cl_2$ and $H_2S$ has been shown in Fig. 9 (right) and the extracted parameters have been plotted in Fig. 10. Both $Cl_2$ (oxidizer) and $H_2S$ (reducer) when exposed, replace adsorbed air indicating stronger interaction and shortening of bond lengths Zn-O3 and Zn-O1 than in the air [Fig. 10 (a) and (b)]. Therefore presence of ambient gases leads to expansion of lattice and average anion-cation bond lengths in NPs. It explains the similar trend observed in Zn-O1 and Zn-O3 with size for small spherical particles (Fig. 6). It should be mentioned that only Type I and Type II samples exhibit significant change in characteristics for different ambient gases due to large area of interaction per unit volume.

A uniform array of dipoles creates surface charge which exerts net electric field in a direction opposite to the direction of the dipoles. Further, the unsaturated dangling bonds at the surface are generally compensated by adsorption of reactive molecules such as $H_2O$ and oxygen [34] and the moment associated with dangling bonds is weaker than Zn-O bonds in the bulk. Thus external factors like adsorbed species may induce relaxation and reconstruction of surface which produces repulsive force (negative pressure) resulting in stretching of anion-cation bond lengths. Hence the average bond length is higher for smaller NPs having high surface to volume ratio. As shown in Fig. 10 (d), number of nearest neighbour Zn atoms around a centre Zn atom [coordination number of the Zn-Zn shell] decrease considerably for smaller NPs from the model compound value 6. It results increase in Debye-Waller factor significantly [Fig. 10 (c)] indicating drastic break down of the ZnO structure as the particle size decreases due to which the 3rd coordination shell is much more affected than the first two oxygen shells [36,37]. The effect is more pronounced for the samples exposed to $Cl_2$ atmosphere while exposure to $H_2S$ somehow retards the change in structure. $Cl_2$ occupies an oxygen vacancy site thereby imparting stronger interaction with the Zn atom. On the other hand H$_2$S is known for donating an electron keeping intact the chemical nature of ZnO [35].

## 5. Conclusions

In conclusion, size and shape dependent structural parameters of ZnO NS have been investigated experimentally as well as by theoretical modelling. Flat faces of NRs appear, as the size of the nanosphere gradually increases due to relaxation of built up strain. The change in shape from sphere to rod is associated with contraction of lattice constant *a* and Zn-O1 bond along c-axis but stretching of off *c*-axis Zn-O3 bonds which may explain many contradictory results on structural properties of ZnO based materials. With increase in size, *c* decreases monotonously and cell volume contracts for smaller NPs but remains almost constant for bigger NRs. The large surface dipole moment at the side wall predicted for ZnO NRs may be responsible for increase in Zn-O3 and *a* seen after shape transition. The lattice expansion seen for smaller spherical NPs is explained in terms of reduced electrostatic forces due to surface dipoles modified by the presence of ambient gases.

**Acknowledgements**

MG is thankful to Dr. Sudhindra Rayaprol and Dr. Som Datta Kaushik for their help in analyzing the X-ray diffraction results and also acknowledges the support received from Dr. Ajay Singh during EXAFS measurement. X-ray diffraction data in part were collected from S. N. Bose National Centre for Basic Sciences, Kolkata. DK would like to acknowledge the Indo-EU project MONAMI and BARC supercomputing facility for theoretical calculation.

**Figure Captions**

**Fig. 1.** (color online) Aspect ratio (AR) vs. size plot of nearly spherical (Type I), rods (Type II) and long rods (Type III) of ZnO nanostructures exhibit tilted Y-shape characteristic.

**Fig. 2.** (color online) Full-Prof Fitting of X-ray Diffraction data from ZnO NPs of various shapes and sizes.

**Fig. 3.** (color online) Important structural parameters of hexagonal ZnO (left). The theoretical cluster systems S, C1, C2 and C3 having ARs 1, 1.7, 4.3 and 7.1 respectively.

**Fig. 4.** (color online) Internal parameter (u) exhibit approximately linear dependence with the aspect ratio shown in logarithmic scale. Corresponding sizes are indicated.

**Fig. 5.** (color online) Anion-cation bond lengths (top) along c-axis [Zn-O1 (co-ordination number N=1)] and (bottom) off c-axis [Zn-O3 (N=3)] as a function of internal parameter ($u$) (left) and aspect ratio (right). X-axis of right panel is shown in logarithmic scale.

**Fig. 6.** (color online) Size dependence of Zn-O bond lengths show tilted Y-shape characteristic. AR is indicated near data points.

**Fig. 7.** (color online) Lattice parameters *a*, *c*, and *V* vs. size. AR is indicated near data points.

**FIG. 8.** (color online) Micro-strain as a function of particle size and aspect ratio (inset bottom).Variation of bond angle between two Zn-O3 bond (α) with size (inset top).

**Fig. 9.** (color online) Experimental EXAFS [μ(E) vs E ] and absorption function [χ(R)] of 10 nm NPs after the exposure of different gases.

**Fig. 10.** (color online) Variation of (a) Zn-O3, (b) Zn-O1 (c) Debye Waller factor and (d) Zn coordination number of the Zn−Zn shell with particle size when exposed to different gases.